\documentstyle[aps,preprint,tighten,floats,axodraw,epsf]{revtex}

\newcommand{\Tr}{\mathrm{Tr}}
\newcommand{\s}{\! \not \!}

\begin{document}

\draft

\title{Leading logarithm calculation of the\\
$e^+ e^- \rightarrow e^+\, \nu_e
\, \bar{u}\, d$ cross section}

\author{Parvez Anandam}

\address{Institute of Theoretical Science\\
University of Oregon, Eugene, OR 97403}

\date{9 June 2000}

\maketitle

\begin{abstract}
We analytically evaluate in the leading logarithm approximation the
differential cross section for $e^+ e^- \rightarrow e^+\, \nu_e \,
\bar{u}\, d$. We compare our order $\alpha^4 \alpha_s^0$ leading-log
result to the order $\alpha^4 \alpha_s^0$ exact result
obtained from the GRC4F Monte Carlo program. Finally we use the
Gl\"uck, Reya, Schienbien distribution of partons in a virtual
photon, which incorporates both evolution and nonperturbative strong
interaction contributions, to obtain better estimates of the
differential cross section.
\end{abstract}


\section{Introduction}

Events at a high energy $e^+e^-$ collider in which only hadrons are
seen in the final state and the net momemtum of the hadrons transverse
to the beam axis is large could be signatures for physics beyond the
standard model~\cite{OPAL}. However, these events could also arise
from the simple standard model process depicted in
Fig.~\ref{process}. Here the positron emits a slightly virtual photon
and escapes down the beam pipe. The hadrons recoil against a high
$P_T$ neutrino.

The cross section for background processes that arise from graphs like
that shown in Fig.~\ref{process} can be computed exactly to order
$\alpha^4\alpha_s^0$ by using the Monte Carlo code
GRC4F~\cite{GRC}. However, it is useful to have an approximate
analytical calculation of the cross section as a check, if the
calculation is simple enough to illuminate the basic physics. In this
paper, we provide such a calculation based on the leading logarithm
approximation in which the photon carries small transverse momentum $p_T$
while the quark carries much larger transverse momentum $k_T$: $m_e^2
\ll p_T^2 \ll k_T^2 \ll M_W^2$. We compare our order
$\alpha^4\alpha_s^0$ leading-log result with the order
$\alpha^4\alpha_s^0$ exact result.

We also go beyond the $\alpha^4\alpha_s^0$ leading-log result by
incorporating two strong interaction effects: the DGLAP evolution of
parton distribution functions in a virtual photon giving rise to
contributions of order $\sum_N (\alpha_s \ln Q^2/|p^2|)^N$, and the
phenomenological, nonperturbative splitting of the photon into a
quark plus anything. To that end, we use the Gl\"uck, Reya and
Schienbein (GRS) parton distributions in a virtual
photon~\cite{GRS}. We investigate whether these two strong interaction
effects are big enough to substantially affect the cross section.

\begin{figure}[ht]
\begin{center}
\begin{picture}(200,240)(0,0)
\ArrowLine(60,50)(10,50)  \Text(35,60)[]{$P$} \Text(15,45)[]{$e^+$}
\ArrowLine(85,6.70)(60,50) \Text(80,33)[]{$P'$} \Text(75,10)[]{$e^+$}
\Photon(60,50)(85,93.30){-3}{4.5}  \Text(65,75)[]{$p$}  \Text(80,65)[]{$\gamma$}
\ArrowLine(135,93.30)(85,93.30)   \Text(110, 100)[]{$k''$} \Text(130,85)[]{$\bar{u},d$}
\ArrowLine(85,93.30)(85,143.30)  \Text(77,118)[]{$k$} \Text(90,118)[l]{$u,\bar{d}$}
\ArrowLine(85,143.30)(135,143.30) \Text(110,152)[]{$k'$}  \Text(130,135)[]{$d,\bar{u}$}
\Photon(85,143.30)(60,186.60){-3}{4.5} \Text(65,160)[]{$q$}  \Text(88,170)[]{$W$}
\ArrowLine(10,186.60)(60,186.60) \Text(35,195)[]{$l$}  \Text(15,180)[]{$e^-$}
\ArrowLine(60,186.60)(85,229.90) \Text(65,210)[]{$l'$} \Text(90,225)[]{$\nu_e$}
\end{picture}
\end{center}
\caption{Charged current deep inelastic scattering $e^+ e^-
\rightarrow e^+ \nu_e \, \bar{u}\, d$.}. 
\label{process}
\end{figure}

\section{Framework}

In this section, we set the framework for the approximations we will
make and the momentum cuts we will use. We first present notation that
will prove useful later on in the calculation. We will use light-cone
coordinates~\cite{Soper} to describe momentum four-vectors $p^\mu$,
\begin{eqnarray}  
p^\pm = \frac{ p^0 \pm p^3}{\sqrt{2}}, \ \ \ p_T^j = (p^1,
p^2) .
\end{eqnarray} 
This set of coordinates is very convenient for symmetric collisions
because one of the incoming particles has almost exclusively plus
momentum and negligible minus and transverse momentum, whereas the
other has almost exclusively minus momentum.

There are constraints on the momenta of the photon and the quark
coupling to the W boson. As the transverse momentum squared $p_T^2$ of
the photon falls below the mass squared $m_e^2$ of the positron, the
matrix element squared falls off sharply. We can thus take $m_e^2$ to
be an effective lower limit of integration for $p_T^2$. Its upper
limit is the squared veto momentum of positron detection, $M^2$. This
is an experimental limit and will depend on the exact configuration of
the detector. If the scattered positron has an angle $\theta$ greater
than some limit $\theta_0$, it will be detected. Thus we demand that
$|\theta| < \theta_0$. In our approximate calculation, this amounts to
requiring that $p_T < M = \theta_0 \sqrt{s}/2$.

\begin{figure}[ht]
\begin{center}
\begin{picture}(250,250)(0,0)
\Line(10,10)(210,10)
\Line(10,10)(10,210)
\DashLine(10,10)(210,210){3}   
\DashLine(30,10)(30,210){3}   \Text(30,0)[]{$\ln m_e^2$}
\DashLine(10,70)(100,70){3}   \Text(-10,70)[]{$\ln m_{\rho}^2$}
\DashLine(10,180)(210,180){3}  \Text(-10,180)[]{$\ln Q^2$}
\DashLine(140,120)(140,200){3} \Line(140,10)(140,13) \Text(140,0)[]{$\ln M^2$}
\Line(10,30)(13,30) \Text(-10,30)[]{$\ln m_e^2$}
\Line(10,140)(13,140) \Text(-10,140)[]{$\ln M^2$}
\Line(70,10)(70,13) \Text(70,0)[]{$\ln m_\rho^2$}
\Line(180,10)(180,13) \Text(180,0)[]{$\ln Q^2$}
\Line(30,180)(140,180)
\Line(30,180)(30,70)
\Line(30,70)(70,70)
\Line(70,70)(140,140)
\Line(140,140)(140,180)
\Text(10,220)[]{$\ln k_T^2$}
\Text(230,10)[]{$\ln p_T^2$}
\end{picture}
\end{center}
\caption{Momentum cuts.}
\label{momentumcuts}
\end{figure}
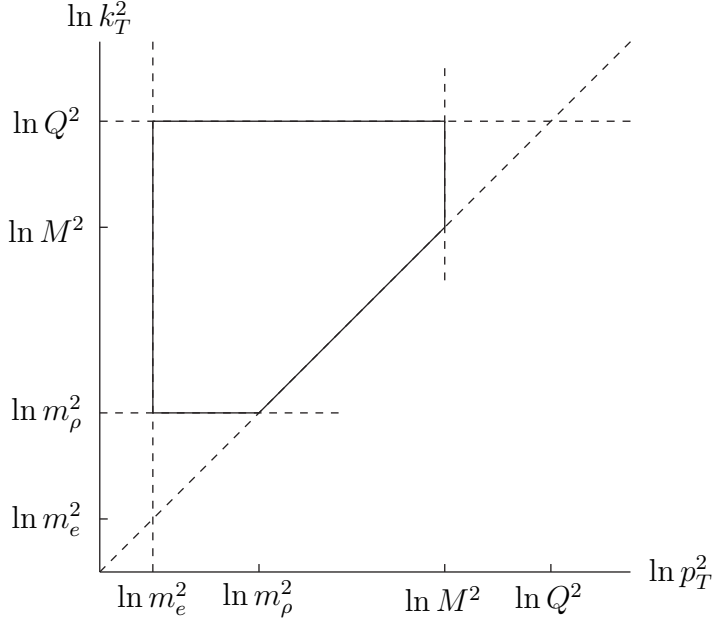

A reasonable lower cut-off on the quark transverse momentum squared
$k_T^2$ is the greater of $m_\rho^2$, the mass squared of the $\rho$
meson (or indeed some other meson such as the $\pi$), and $p_T^2$,
i.e.~$k_T^2 > \max(m_\rho^2, p_T^2)$. For $k_T^2 < m_\rho^2$, the
distribution function of quarks in a photon is nonperturbative. In
the first part of this investigation, we have set this
nonperturbative contribution to zero. The approximate upper limit on
$k_T^2$ is the virtuality of the W boson, $Q^2$, which is fixed by the
transverse momentum of the neutrino. A plot of these constraints,
Fig.~\ref{momentumcuts}, makes it easy to visualize the phase-space
under consideration.

The approximation we make is to calculate the cross section by
including the region in the middle of the polygon in
Fig.~\ref{momentumcuts}, without worrying about what happens at its
edges. This is reasonable because most events will lie in this region,
\begin{eqnarray}
m_e^2 \ll p_T^2 \ll k_T^2 \ll Q^2 .
\end{eqnarray}
This approximation is called the leading-log approximation, for
reasons that will become apparent later on. We can split our
calculation into three steps, in which the hardness increases at each
step.

The first step is to assume that the transverse momentum $p_T$ of the
photon is much smaller than the transverse momentum $k_T$ of the quark
that couples the photon to the W boson (Graph A,
Fig.~\ref{graphA}). We can then calculate the piece of Fig.~\ref{process}
dealing with the emission of the photon from the positron.
The second step is to assume that the transverse momentum $k_T$ of the
quark is much smaller than the transverse momentum $q_T$ of the W
boson (Graph B, Fig.~\ref{graphB}). We then calculate the
piece of Fig.~\ref{process} in which the photon splits into a
quark-antiquark pair.
The third and final step involves no further approximations and is a
direct calculation of the cross section for charged current deep
inelastic scattering (Graph C, Fig.~\ref{graphC}).

The general form of the cross section for the process in
Fig.~\ref{process} is the following:
\begin{eqnarray}
d\sigma & = & \frac{1}{2 P^0 2 l^0 2}|{\mathcal M}_{\mathrm{A}}|^2
\frac{d^2P'_T d P'^+}{(2\pi)^3 2 P'^{+}} \frac{d^2l'_T d
l'^+}{(2\pi)^3 2 l'^+} \frac{d^2k'_T d k'^+}{(2\pi)^3 2 k'^+}
\frac{d^2k''_T d k''^+}{(2\pi)^3 2 k''^+} \nonumber \\ & & \times
(2\pi)^4 \delta^4(P^\mu + l^\mu - P'^{\mu} - l'^{\mu} - k'^{\mu} -
k''^{\mu}) .
\end{eqnarray}
We are now ready to calculate the total matrix element squared.

\section{Graph A: Calculation of $P_{\gamma/\lowercase{e}}$}

As just discussed, we will calculate the total matrix element squared
$|{\mathcal M}_{\mathrm{A}}|^2$ in three successive refinements. First
we concentrate on the part of the diagram involving the photon (Graph
A, Fig.~\ref{graphA}), which will be the starting point. This is the
first step in our three-step calculation, as we go from the least hard
to the most hard part of the diagram in
Fig.~\ref{process}. $|{\mathcal M}_{\mathrm{A}}|^2$ will contain a
factor $|{\mathcal M}_{\mathrm{B}}|^2$ that represents the next
hardest contribution, while $|{\mathcal M}_{\mathrm{B}}|^2$ will
contain the very hardest factor $|{\mathcal M}_{\mathrm{C}}|^2$.

\begin{figure}[ht]
\begin{center}
\begin{picture}(230,200)(0,0)
\ArrowLine(60,50)(10,50)  \Text(35,60)[]{$P$}
\ArrowLine(85,6.70)(60,50) \Text(80,33)[]{$P'$}
\Photon(60,50)(85,93.30){-3}{4.5}  \Text(65,75)[]{$p$}
\Oval(115,110.62)(22.91,45.83)(0)  \Text(116,110.62)[]{${\mathcal M}_{\mathrm{B}}$}
\Photon(145,93.30)(170,50){-3}{4.5}
\ArrowLine(220,50)(170,50)
\ArrowLine(170,50)(145,6.70)
\ArrowLine(60,171.24)(85,127.94) \Text(65,150)[]{$l$}
\ArrowLine(145,127.94)(170,171.24)
\DashLine(115,0)(115,100){3}
\DashLine(115,120)(115,180){3}
\end{picture}
\end{center}
\caption{Graph A.}
\label{graphA}
\end{figure}

We calculate the probability $P_{\gamma/e}$ of finding a photon in a
positron by evaluating Graph~A (Fig.~\ref{graphA}). Averaging over the
initial spins of the positron, we find the total matrix element
squared to be
\begin{eqnarray}
|{\mathcal M}_A|^2 = \frac{e^2}{(p^2)^2} \frac{1}{2} \sum_{i,j}
\Tr\left\{\s{P}\s{\epsilon}(p,j)\s{P}'\s{\epsilon}(p,i)\right\}\delta^{ij}
|{\mathcal M}_B|^2 .
\end{eqnarray}
Here we neglect the positron mass compared to $p_T$. $|{\mathcal
M}_B|^2$ is averaged over the photon polarizations and contains the
next hardest part which we calculate later on. $\epsilon(p,i)$ is the
polarization vector in the physical $A^+=0$ gauge of the photon with
momentum $p^\mu$ and polarization index $i$.

We work in the c.m.\ frame of the electron and positron, with the
positron carrying only plus momentum and the electron carrying only
minus momentum. In this frame the momenta in our graphs have
components $v^\nu = (v^+, v^-, \vec{v}_T)$ given by
\begin{eqnarray}
P^\mu & = & (P^+,0,\vec{0}_T), \nonumber\\ 
l^\mu & = & (0,l^-,\vec{0}_T), \nonumber\\ 
P'^{\mu} & = & \left( (1-\xi)P^+, 
\frac{p_T^2}{2(1-\xi)P^+}, -\vec{p}_T\right), \nonumber\\ 
p^\mu & = & \left(\xi P^+, \frac{-p_T^2}{2(1-\xi)P^+},
\vec{p}_T\right).
\end{eqnarray}
Now, $\epsilon^\mu (p,i)$ is
\begin{eqnarray}
\epsilon^+(p,i) = 0, \ \ \ \epsilon^-(p,i) = p_T \cdot \epsilon_T/p^+
, \ \ \ \epsilon^j(p,i)=\delta_i^j .
\end{eqnarray}  
Using $\sum_i \epsilon^2(p,i) = -2$, we
have
\begin{eqnarray}
|{\mathcal M}_{\mathrm{A}}|^2 = \frac{4(2\pi)\alpha}{p_T^2}
\frac{(1-\xi)}{\xi} \left\{ \frac{1+(1-\xi)^2}{\xi}\right\} |{\mathcal
M}_{\mathrm{B}}|^2 .
\end{eqnarray}
We recognize the standard splitting function $P_{\gamma/e}$,
\begin{eqnarray}
P_{\gamma/e}(\xi) = \frac{1+(1-\xi)^2}{\xi} .
\end{eqnarray}

Define $p^\mu \approx \tilde{p}^\mu \equiv (\xi P^+, 0, \vec{0}_T)$. Since the process is very
hard, i.e.~$p_T \ll k_T$, the minus components and transverse
components of $p^\mu$ are small compared to the plus components. In this
approximation,
\begin{eqnarray}
(2\pi)^4 \delta^4(P^\mu + l^\mu - P'^{\mu} - l'^{\mu} - k'^{\mu} -
k''^{\mu}) \approx (2\pi)^4 \delta^4(\tilde{p}^\mu + l^\mu - l'^{\mu}
- k'^{\mu} - k''^{\mu}).
\end{eqnarray}
Then the cross section can be reexpressed to depend on $p_T$ and
$\xi$,
\begin{eqnarray}
d\sigma & = & \frac{2\alpha d^2p_Td\xi}{(2\pi)^2
p_T^2}P_{\gamma/e}(\xi) \frac{1}{2\tilde{p}^0 2l^0 2}|{\mathcal
M}_{\mathrm{B}}|^2 \frac{d^2l'_T d l'^+}{(2\pi)^3 2 l'^+}
\frac{d^2k'_T d k'^+}{(2\pi)^3 2 k'^+} \frac{d^2k''_T d
k''^+}{(2\pi)^3 2 k''^+} \nonumber \\ 
& & \times (2\pi)^4 \delta^4(\tilde{p}^\mu + l^\mu - l'^{\mu} 
- k'^{\mu} - k''^{\mu}) .
\end{eqnarray}

\section{Graph B: Calculation of $P_{\lowercase{q}/\gamma}$}

We now calculate the probability of finding a quark in a photon. This
will give us $|{\mathcal M}_{\mathrm{B}}|^2$, in terms of the final
hard matrix element squared $|{\mathcal M}_{\mathrm{C}}|^2$. There is
also a graph $\mathrm{B}'$ in which the quark is replaced by an
antiquark. In this case, everything is the same except the hard matrix
element $|{\mathcal{M}}_{\mathrm{C}}|^2$ is replaced by a different
matrix element $|{\mathcal{M}}_{\mathrm{C}'}|^2$.

\begin{figure}
\begin{center}
\begin{picture}(145,200)(10,10)
\Photon(-0.36,17.94)(35,53.30){-3}{4.5}  \Text(5,37)[]{$p$}
\ArrowLine(35,53.30)(60,96.60)  \Text(37,75)[]{$k$}
\BCirc(72.5,118.25){25} \Text(72.5,118.25)[]{${\mathcal M}_{\mathrm{C}}$}
\ArrowLine(85,96.60)(110,53.30)
\Photon(110,53.30)(145.36,17.94){-3}{4.5}
\ArrowLine(110,53.30)(35,53.30) \Text(75,62)[]{$k''$}
\ArrowLine(35,183.20)(60,139.9) \Text(37,162)[]{$l$}
\ArrowLine(85,139.9)(110,183.20)
\DashLine(72.5,10)(72.5,47){3}
\DashLine(72.5,70)(72.5,108){3}
\DashLine(72.5,125)(72.5,190){3}
\end{picture}
\end{center}
\caption{Graph B.}
\label{graphB}
\end{figure}

Summing over colors and averaging over spins, the matrix element
squared for Graph~B (Fig.~\ref{graphB}) is
\begin{eqnarray}
|{\mathcal M}_{\mathrm{B}}|^2 = N_C \frac{e_q^2 e^2}{(k^2)^2}
\frac{1}{2} \sum_i \Tr \left\{{\mathcal
H}\s{k}\s{\epsilon}(\tilde{p},i)\s{k}''\s{\epsilon}(\tilde{p},i)\s{k} \right\} .
\end{eqnarray}
Again, the mass of the quark is negligible compared to $k_T$. The
matrix ${\mathcal H}$ contains the final contribution that we will
evaluate using Graph~C (Fig.~\ref{graphC}) later on. For now, we leave
it as it is. Let
$\s{L}=\s{k}\s{\epsilon}(\tilde{p},i)\s{k}''\s{\epsilon}(\tilde{p},i)\s{k}$. Since the
partons involved are going much faster in the plus direction compared
to all other directions, we can make the approximation $\s{L} \approx
L^+\gamma^-$. Then,
\begin{eqnarray}
|{\mathcal M}_{\mathrm{B}}|^2 = N_C \frac{e_q^2 e^2}{(k^2)^2}
\frac{1}{2} \sum_i \Tr \left\{ {\mathcal H}\gamma^- \right\} L^+ .
\end{eqnarray}
Let $\tilde{k}^\mu = (k^+,0,\vec{0}_T)$. We use it to rewrite the
above,
\begin{eqnarray}
|{\mathcal M}_{\mathrm{B}}|^2 = N_C \frac{e_q^2 e^2}{(k^2)^2}
\frac{1}{2} \sum_i \Tr \left\{ {\mathcal H}\s{\tilde{k}} \right\}
\frac{L^+}{k^+} , \nonumber \\ 
L^+ = \frac{1}{4} \Tr \left\{ \gamma^+
L \right\} = \frac{1}{4} \Tr \left\{
\gamma^+\s{k}\s{\epsilon(\tilde{p},i)}\s{k}''\s{\epsilon(\tilde{p},i)}\s{k} \right\} .
\end{eqnarray}
The hard scattering matrix element squared, which we leave unevaluated
for now, is $|{\mathcal M}_{\mathrm{C}}|^2 = \frac{1}{2} \Tr \left\{
{\mathcal H}\s{\tilde{k}} \right\}$, and the matrix element squared is
in the simple form
\begin{eqnarray}
|{\mathcal M}_{\mathrm{B}}|^2 = N_C \frac{e_q^2 e^2}{(k^2)^2} \sum_i
\frac{1}{4} \frac{1}{k^+} \Tr \left\{
\gamma^+\s{k}\s{\epsilon(\tilde{p},i)}\s{k}''\s{\epsilon(\tilde{p},i)}\s{k} \right\}
|{\mathcal M}_{\mathrm{C}}|^2 .
\end{eqnarray}

In the same c.m. frame as earlier, but now with the
approximation $p^\mu \rightarrow \tilde{p}^\mu$ described above,
\begin{eqnarray}
p^\mu & = & (p^+,0,\vec{0}_T) , \nonumber\\ 
k^\mu & = & \left( z p^+, \frac{-k_T^2}{2(1-z)p^+}, \vec{k}_T \right) , 
\nonumber\\ 
k''^{\mu} & = & \left( (1-z)p^+,
\frac{k_T^2}{2(1-z)p^+}, -\vec{k}_T\right) .
\end{eqnarray}
After doing the trace algebra, the matrix element squared is
\begin{eqnarray}
|{\mathcal M}_{\mathrm{B}}|^2 = \frac{4(2\pi)\alpha}{k_T^2}
\frac{(1-z)}{z} e_q^2 N_C \left\{ z^2+(1-z)^2\right\} |{\mathcal
M}_{\mathrm{C}}|^2.
\end{eqnarray}
We recognize the standard splitting function $P_{q/\gamma}$,
\begin{eqnarray}
P_{q/\gamma}(z) = N_C\left\{ z^2+(1-z)^2 \right\} .
\end{eqnarray}

In the same approximation as before, the minus and transverse
components being small compared to the plus components, the delta
function in the cross section can be expressed as
\begin{eqnarray}
(2\pi)^4 \delta^4(\tilde{p}^\mu + l^\mu - l'^{\mu} - k'^{\mu} -
k''^{\mu}) \approx (2\pi)^4 \delta^4(\tilde{k}^\mu + l^\mu - l'^{\mu}
- k'^{\mu}) .
\end{eqnarray}
Let $x=k^+/P^+$, which implies $k^+=(x/\xi)p^+$ and $z=x/\xi$. Then,
using the result we obtained for $|{\mathcal M}_{\mathrm{B}}|^2$, the
cross section can be written as
\begin{eqnarray}
d\sigma & = & \frac{4 \alpha^2 e_q^2}{(2\pi)^4} \frac{d\xi}{\xi} dx
P_{\gamma/e}(\xi) P_{q/\gamma}(x/\xi) \frac{d^2p_T}{p_T^2}
\frac{d^2k_T}{k_T^2} \nonumber \\ & & \times \frac{1}{2\tilde{k}^0
2l^0 2} |{\mathcal M}_{\mathrm{C}}|^2 \frac{d^2l'_T d l'^+}{(2\pi)^3 2
l'^+} \frac{d^2k'_T d k'^+}{(2\pi)^3 2 k'^+} (2\pi)^4
\delta^4(\tilde{k}^\mu + l^\mu - l'^{\mu} - k'^{\mu}) ,
\end{eqnarray}
which makes it obvious that the cross section is the convolution of a
hard-scattering cross section with a piece that we identify as the
distribution function of a quark in a positron.

A brief description of the required distribution functions is
relevant. Graph~A (Fig.~\ref{graphA}) involves the distribution
function of the photon in the positron, and Graph~B
(Fig.~\ref{graphB}) the distribution function of a quark in a
photon. As we combine the two graphs, we will require the distribution
function of a quark in a positron. We will perform the trivial angular
integral to write $d^2 p_T = \pi d p_T^2$. We now define the
distribution functions described,
\begin{eqnarray}
f_{\gamma/e}(x) & = & \frac{\alpha}{2\pi} \int_{m_e^2}^{M^2} \frac{d
p_T^2}{p_T^2} P_{\gamma/e}(x) , \nonumber\\ f_{q/\gamma}(x,Q^2,p_T^2)
& = & \frac{\alpha e_q^2}{2\pi}\int_{\max(p_T^2,m_\rho^2)}^{Q^2}
\frac{d k_T^2}{k_T^2} P_{q/\gamma}(x) , \nonumber\\ f_{q/e}(x, Q^2) &
= & \frac{\alpha^2 e_q^2}{(2\pi)^2} \int_x^1 \frac{d\xi}{\xi}
P_{\gamma/e}(\xi) P_{q/\gamma}(x/\xi) \int_{m_e^2}^{M^2} \frac{d
p_T^2}{p_T^2}
\int_{\max(p_T^2, m_\rho^2)}^{Q^2} \frac{d k_T^2}{k_T^2} .
\end{eqnarray}
We first evaluate the convolution of the splitting functions,
\begin{eqnarray}
\int_x^1 \frac{d\xi}{\xi} P_{\gamma/e}(\xi) P_{q/\gamma}(x/\xi) =  
N_C \left( (1-x) + 2(1+x)\ln x + \frac{4}{3} \frac{(1-x^3)}{x}
\right) .
\end{eqnarray}
Next, we compute the momentum integral, either by inspection of
Fig.~\ref{momentumcuts}, or by explicit calculation,
\begin{eqnarray}
\int_{m_e^2}^{M^2} \frac{d p_T^2}{p_T^2}
\int_{\max(p_T^2,m_\rho^2)}^{Q^2} \frac{d k_T^2}{k_T^2} & = & \ln
\frac{Q^2}{m_\rho^2} \ln \frac{m_\rho^2}{m_e^2} + \ln \frac{Q^2}{M^2}
\ln \frac{M^2}{m_\rho^2} + \frac{1}{2} \ln^2 \frac{M^2}{m_\rho^2} .
\end{eqnarray}
Therefore, the distribution function of a quark in a positron (or,
equivalently, in an electron) is
\begin{eqnarray}
f_{q/e}(x, Q^2) = \frac{\alpha^2 e_q^2}{(2\pi)^2} N_C \left( (1-x) +
2(1+x)\ln x + \frac{4}{3} \frac{(1-x^3)}{x} \right) \nonumber \\
\times \left( \ln \frac{Q^2}{m_\rho^2} \ln \frac{m_\rho^2}{m_e^2} +
\ln \frac{Q^2}{M^2} \ln \frac{M^2}{m_\rho^2} + \frac{1}{2} \ln^2
\frac{M^2}{m_\rho^2} \right) .
\end{eqnarray}

If we use phenomenological distribution functions $f_{q/\gamma}$ of
quarks in virtual photons (e.g. GRS distribution functions), we can
evaluate $f_{q/e}$ as
\begin{eqnarray}
f_{q/e}(x,Q^2) & = & \frac{\alpha}{2\pi}\int_x^1\frac{d \xi}{\xi}
\int_{m_e^2}^{M^2} \frac{d p_T^2}{p_T^2} P_{\gamma/e}(x/\xi)
f_{q/\gamma}(\xi, Q^2, p_T^2) .
\end{eqnarray}
We will use this convolution in a following section when using the GRS
distribution functions.

We label the hard scattering cross section $d \hat{\sigma}$,
\begin{eqnarray}
d\hat{\sigma} = \frac{1}{2\tilde{k}^0 2l^0 2} |{\mathcal
M}_{\mathrm{C}}|^2 \frac{d^2l'_T d l'^+}{(2\pi)^3 2 l'^+}
\frac{d^2k'_T d k'^+}{(2\pi)^3 2 k'^+} (2\pi)^4 \delta^4(\tilde{k}^\mu
+ l^\mu - l'^{\mu} - k'^{\mu}) .
\end{eqnarray} 
The cross section $d\sigma$ is a convolution of the distribution
function with the hard scattering cross section
\begin{eqnarray}
d\sigma = \int dx \, f_{q/e}(x) \, d\hat{\sigma}.
\end{eqnarray}

\section{Graph C: Charged current deep inelastic scattering}

\begin{figure}[ht]
\begin{center}
\begin{tabular}{cc}
\hspace{0.5in}
\begin{picture}(100,140)(0,0)
\ArrowLine(10,10)(45.36,45.36)   \Text(26,35)[]{$k$}
\ArrowLine(45.36,45.36)(80.72,10)   \Text(62,35)[l]{$k'=k+q$}
\Photon(45.36,45.36)(45.36,95.36){-3}{4.5}    \Text(35,70)[]{$q$} \Text(57,70)[]{$W$}
\ArrowLine(10,130.72)(45.36,95.36)   \Text(30,120)[]{$l$}
\ArrowLine(45.36,95.36)(80.72,130.72) \Text(62,120)[]{$l'$}
\end{picture}
\hspace{0.5in}
&
\hspace{0.5in}
\begin{picture}(100,140)(0,0)
\ArrowLine(45.36,45.36)(10,10)   \Text(26,35)[]{$k$}
\ArrowLine(80.72,10)(45.36,45.36)   \Text(62,35)[l]{$k'=k+q$}
\Photon(45.36,45.36)(45.36,95.36){-3}{4.5}    \Text(35,70)[]{$q$} \Text(57,70)[]{$W$}
\ArrowLine(10,130.72)(45.36,95.36)   \Text(30,120)[]{$l$}
\ArrowLine(45.36,95.36)(80.72,130.72) \Text(62,120)[]{$l'$}
\end{picture}
\hspace{0.5in}
\\
Graph C & Graph $\mathrm{C}'$
\end{tabular}
\end{center}
\caption{Charged current deep inelastic scattering.}
\label{graphC}
\end{figure}
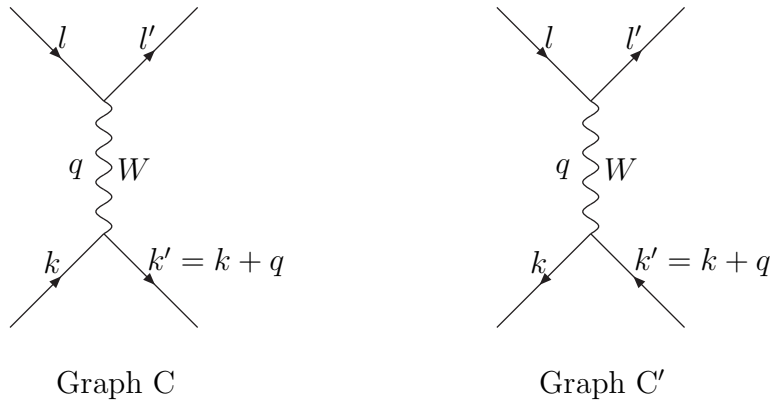

The third and final part of the calculation is to compute the charged
current deep inelastic scattering cross section, Graph~C
(Fig.~\ref{graphC}). This final calculation is quite standard and
involves no further approximations. It is a direct calculation of the
Feynman diagrams in Fig.~\ref{graphC}. It involves the W boson
coupling constant $g_W=\sqrt{4\pi\alpha}/\sin \theta_W$ and the
element of the quark mixing matrix $V_{ud}$ that relates the up quark
to the down quark.

Averaging over initial spins, the matrix element squared for Graph~C
(Fig.~\ref{graphC}) is
\begin{eqnarray}
|{\mathcal M}_{\mathrm{C}}|^2 & = & \frac{2 \hat{s}^2 (2 \pi)^2
\alpha^2 |V_{ud}|^2}{\sin^4 \theta_W (Q^2 + M_W^2)^2},
\end{eqnarray}
where $\hat{s}$ is the c.m.\ energy squared of the electron-quark system. 

The calculation of the matrix element squared for an incoming
antiquark instead of a quark is nearly identical, and is depicted in
Graph~$\mathrm{C}'$, Fig.~\ref{graphC}. The matrix element squared is
\begin{eqnarray}
|{\mathcal M}_{\mathrm{C}'}|^2 & = & \frac{2 \hat{s}^2 (2 \pi)^2
\alpha^2 |V_{ud}|^2}{\sin^4 \theta_W (Q^2 + M_W^2)^2} (1-y)^2,
\end{eqnarray}
where $y \equiv q\cdot P / l\cdot P$.

We are now ready to evaluate the hard scattering cross section. We
first rewrite it in a more suitable form,
\begin{eqnarray}
d\hat{\sigma} & = & \frac{1}{2\tilde{k}^0 2l^0 2}
\frac{d^2l'_T}{(2\pi)^2} \frac{dl'^+}{2l'^+} \delta(k'^2)|{\mathcal
M}_{\mathrm{C}}|^2 .
\end{eqnarray} 
Now,
\begin{eqnarray}
k'^2 = (k+q)^2 =  2 x q \cdot P - Q^2 = Q^2(x/x_{bj}-1) ,
\end{eqnarray}
where $x_{bj} \equiv Q^2/2q \cdot P$. We express the momentum $l'$ of
the outgoing lepton in terms of its transverse momentum $l'_T$ and its
pseudo-rapidity $\eta'$,
\begin{eqnarray}
l'^\mu = \left( l'_T e^{\eta'} / \sqrt{2}, l'_T e^{-\eta'} / \sqrt{2},
l'_T \right),
\end{eqnarray}
so $dl'^+ = l'^+ d\eta'$. The azimuthal angle integral is trivial,
$d^2l'_T = \pi dl_T'^2$. In terms of the c.m. energy of the
hard process, $ k^0 l^0 = \hat{s}/4$.

The hard cross section is then
\begin{eqnarray}
d\hat{\sigma} = \frac{1}{16\pi \hat{s}} \frac{x_{bj}}{Q^2}dl_T'^2
d\eta' \delta(x-x_{bj}) |{\mathcal M}_{\mathrm{C}}|^2 .
\end{eqnarray}
Using the squared matrix elements $|{\mathcal{M}}_{\mathrm{C}}|^2$ and $|{\mathcal{M}}_{\mathrm{C}'}|^2$ we calculated previously, and the fact that
$Q^2 = y \hat{s}$, we obtain
\begin{eqnarray}
d\hat{\sigma}_{e^- u \rightarrow \nu_e d} &=& \frac{ \pi \alpha^2
|V_{ud}|^2}{2 \sin^4 \theta_W (Q^2+M_W^2)^2} \frac{x}{y} dl_T'^2
d\eta' \delta(x-x_{bj}) , \nonumber\\ d\hat{\sigma}_{e^- \bar{d}
\rightarrow \nu_e \bar{u}} &=& \frac{ \pi \alpha^2 |V_{ud}|^2}{2 \sin^4 \theta_W
(Q^2+M_W^2)^2} \frac{x}{y} dl_T'^2 d\eta' \delta(x-x_{bj})(1-y)^2 .
\end{eqnarray}

\section{Differential cross sections}

We have computed all the pieces necessary for the calculation of the
cross section,
\begin{eqnarray}
d\sigma = \int dx \left\{ f_{u/e}(x) d\hat{\sigma}_{e^- u \rightarrow
\nu_e d} +f_{\bar{d}/e}(x) d\hat{\sigma}_{e^- \bar{d} \rightarrow
\nu_e \bar{u}} \right\} .
\end{eqnarray}
The result is
\begin{eqnarray}
\frac{d\sigma}{dl_T'^2 d\eta'} = \frac{ \alpha^4 N_C |V_{ud}|^2}{8 \pi \sin^4 \theta_W (Q^2+M_W^2)^2}\left( \frac{1}{2} \ln^2 \frac{M^2}{m_\rho^2} + \ln \frac{M^2}{m_\rho^2} \ln \frac{m_\rho^2}{m_e^2} + \ln \frac{Q^2}{M^2}\ln \frac{M^2}{m_e^2} \right) \nonumber\\
\times \left( x(1-x) + 2x(1+x)\ln x + \frac{4}{3} (1-x^3) \right)
\left( \frac{e_u^2 + e_{\bar{d}}^2(1-y)^2}{y} \right ).
\end{eqnarray}
It is often more convenient to rexpress this in terms of $l_T'$ rather
than $l_T'^2$,
\begin{eqnarray}
\frac{d\sigma}{dl_T' d\eta'} = (2l_T') \frac{d\sigma}{dl_T'^2 d\eta'} .
\end{eqnarray}

The final step is to change variables from $x,y,Q^2$ to $\eta',l_T',s$
where $s$ is the c.m.\ energy squared of the $e^+ e^-$ pair,
not to be confused with the hard scattering c.m.\ energy
squared $\hat{s}$. In the laboratory frame, neglecting $m_e$ in the
change of variables,
\begin{eqnarray}
P^\mu & = & \left(\sqrt{s/2},0,\vec{0}_T \right), \nonumber\\ 
l^\mu & = & \left(0,\sqrt{s/2},\vec{0}_T \right), \nonumber\\ 
l'^\mu & = & \left( \frac{l'_T e^{\eta'}}{\sqrt{2}}, \frac{l'_T
e^{-\eta'}}{\sqrt{2}}, \vec{l}'_T \right).
\end{eqnarray}
The change of variables is
\begin{eqnarray}
Q^2 & = & -(l-l')^2 = e^{\eta'}l_T'\sqrt{s} , \nonumber \\ x & = &
\frac{Q^2}{2 q \cdot P} = \frac{e^{\eta'} l_T'}{\sqrt{s} -
e^{-\eta'}l_T'} , \nonumber \\ y & = & \frac{ q \cdot P}{l \cdot P} =
1-\frac{l_T' e^{-\eta'}}{\sqrt{s}} .
\end{eqnarray}
The physical region is where $x<1$:
\begin{eqnarray}
2 l_T' \cosh \eta' < \sqrt{s} .
\label{physcond}
\end{eqnarray}

We now have a simple approximation to the cross section that can be
compared with the output of Monte Carlo programs. We check our cross
section against GRC4F \cite{GRC}. Later we introduce a refinement not
included in the purely perturbative Monte Carlo calculation. From
GRC4F, we obtain\footnote{We thank Alain Bellerive of the OPAL
collaboration at CERN for providing us with the GRC4F Monte Carlo
events.} the momentum four-vectors of all final state particles for
3666 events of the type we are interested in corresponding to a
luminosity ${\mathrm{L}} = 100\ {\mathrm{pb}}^{-1}$. The c.m. energy
is $\sqrt{s}=189 \ {\mathrm{GeV}}$ and the veto momentum squared was
taken to be $M^2 = 5\ {\mathrm{GeV^2}}$. To be consistent with the
choice made in getting the GRC4F results, we pick the cut-off for the
pertubative treatment of our parton distributions to be twice the pion
mass, i.e. $k_T^2 > \max( (0.28\ \mathrm{GeV})^2, p_T^2)$. We make a
histogram of cross section $d\sigma/d l_T'$ versus transverse momentum
of the neutrino $l_T'$.  Next, we compare this GRC4F result to the
approximate cross section obtained by numerically integrating the
differential cross section we calculated
\begin{eqnarray}
\frac{d\sigma}{d l_T'} =
\int_{-\eta'_{\mathrm{max}}}^{\eta'_{\mathrm{max}}}  d\eta'
\frac{d\sigma}{dl_T' d\eta'} .
\end{eqnarray}
The range of pseudo-rapidity follows directly from the physicality
condition in Eq.~(\ref{physcond}),
\begin{eqnarray}
\eta'_{\mathrm{max}} = \log
\frac{1 + \sqrt{1 - (2 l_T'/\sqrt{s})^2}}{2 l_T'/\sqrt{s}} .
\end{eqnarray}

\begin{figure}[ht]
\vspace{0.2in}
\centerline{\epsfbox{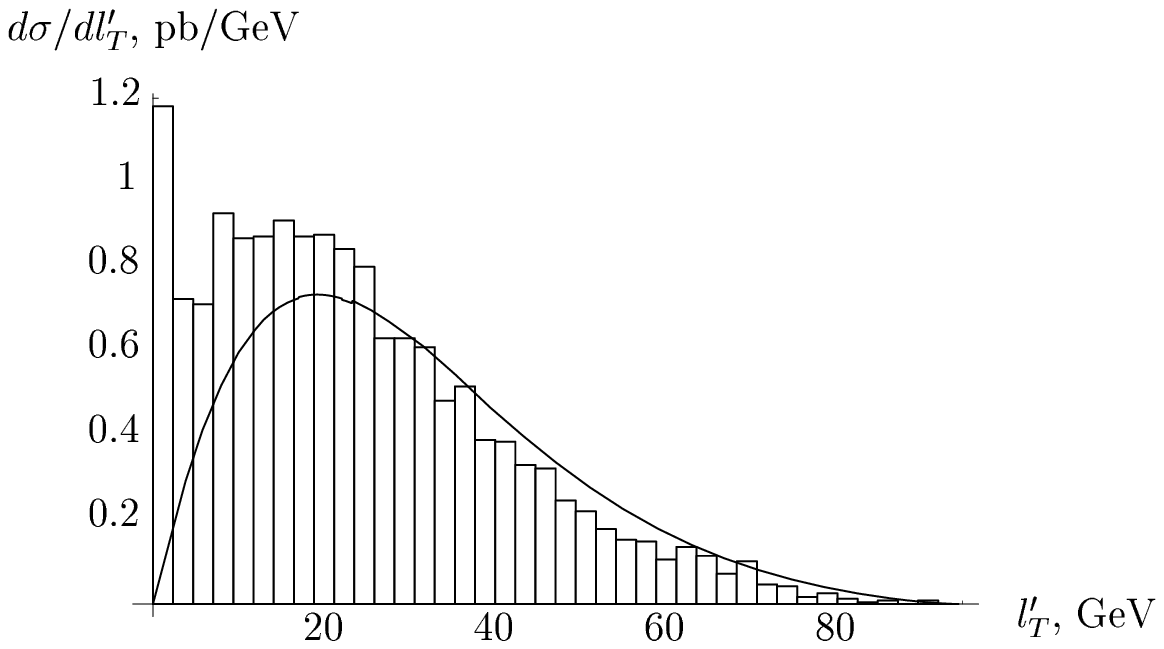}}
\vspace{0.2in}
\caption{Differential cross section $d\sigma/d l'_T$ versus transverse momentum of the neutrino. Our theoretical line is superimposed on results from Monte Carlo simulations using GRC4F with 3666 events.}
\label{plotlt}
\end{figure}

We plot this theoretical curve over the histogram in
Fig.~\ref{plotlt}. Taking into consideration the multiple
approximations we made, we see that our calculated cross section
agrees quite well with the Monte Carlo GRC4F, except at low $l_T'$,
where the leading logarithm approximations used in the approximate
calculation do not apply.

We follow a similar procedure to obtain a graph of $d\sigma/d
\eta'$. We bin the events in pseudorapidity $\eta'$ of
the neutrino, from $\eta'=-10$ to $\eta'=10$, which is a reasonably
large enough range to allow the cross section to fall off sharply at
its edges, and we plot a histogram. Finally, we numerically integrate
the differential cross section we calculated,
\begin{eqnarray}
\frac{d\sigma}{d \eta'} = \int_{0}^{l'_{T{\mathrm{max}}}}  d l_T'
\frac{d\sigma}{dl_T' d\eta'} .
\end{eqnarray}
From the boundary conditions of the physical region, the integration
range of $l_T'$ is from $0$ to $l'_{T{\mathrm{max}}}$,
\begin{eqnarray}
l'_{T{\mathrm{max}}} = \frac{\sqrt{s}}{2 \cosh \eta'} .
\end{eqnarray}  

\begin{figure}[ht]
\vspace{0.2in}
\centerline{\epsfbox{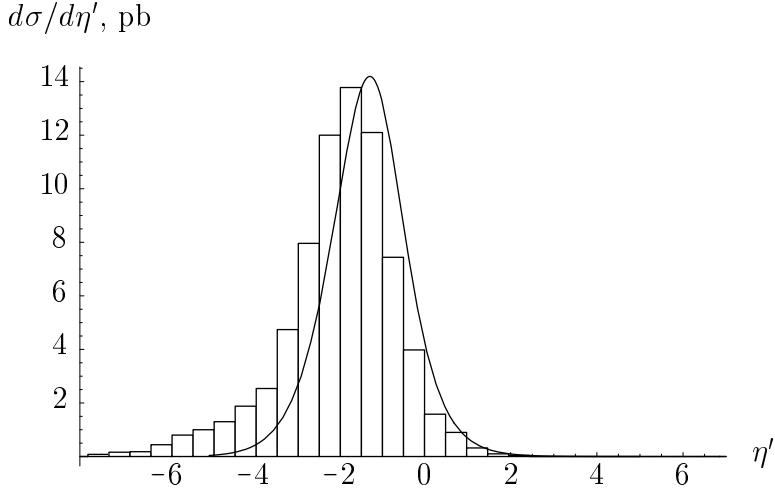}}
\vspace{0.2in}
\caption{Differential cross section $d\sigma/d \eta'$ versus pseudo-rapidity of the neutrino. Our theoretical line is superimposed on results from Monte Carlo simulations using GRC4F with 3666 events.}
\label{ploteta}
\end{figure}

The histogram and the theoretical curve are plotted in
Fig.~\ref{ploteta}. Once again, we see that our approximate
calculation and GRC4F agree except for large negative $\eta'$, where
the approximations we used break down.

As a final check, we compute the total cross section $\sigma$. The
GRC4F cross section is just the total number of events of this process
divided by the luminosity, so $\sigma_{{\mathrm{GRC4F}}}=36.7 \
{\mathrm{pb}}$. We calculate the total cross section from our result
by either integrating $d\sigma/dl_T'$ or $d\sigma/d\eta'$ over their
entire ranges $0<l_T'<\sqrt{s}/2$ and $-\infty<\eta'<\infty$. The
cross section we obtain by either method is $\sigma_{\mathrm{th}}=
30.7 \ {\mathrm{pb}}$. This is in reasonable agreement with the GRC4F
result.

In conclusion, we see that the cross section we calculated
analytically from first principles in the leading log appromation is
not far off from the GRC4F cross section in the region of large $l'_T$
and not too negative $\eta'$, where the momentum transfer from the
electron is large.

\section{Refinement using the GRS parton distributions}

So far, our calculation has been at order $\alpha_s^0$, using
pointlike parton distribution functions of quarks in a photon, without
any evolution. The GRC4F result was exact at order $\alpha^4
\alpha_s^0$, and our approximate calculation was a leading logarithm
calculation at order $\alpha^4 \alpha_s^0$. In this section, we extend
the calculation to include strong interaction effects. We continue to
work in the approximation that virtualities are strongly ordered (the
leading log approximation) since we have seen that this approximation
works well for large $l_T'^2$ and $\eta' \stackrel{>}{\scriptstyle
\sim} -3$. We investigate two strong interaction effects. First,
the emission of gluons from the quark before it is struck can lead to
contributions of order $\sum_N (\alpha_s \ln Q^2/|p^2| )^N$, that tend
to lower the cross section. We can account for these potentially
important contributions by using parton distribution functions
$f_{q/\gamma}$ that have evolved according to the DGLAP evolution
equation. Second, for $|p^2| \stackrel{<}{\scriptstyle \sim}
m_\rho^2$, there can be important nonperturbative hadronic
contributions to $f_{q/\gamma}$.  This ``vector meson dominance''
component can be modelled by using the parton distribution
functions in pions~\cite{GRSpion}, which are determined empirically. We can account
for both these effects by using the recently published
\cite{GRS} parton distribution function for virtual photons by
Gl\"uck, Reya, and Schienbein (GRS).

\begin{figure}[ht]
\vspace{0.2in}
\centerline{\epsfbox{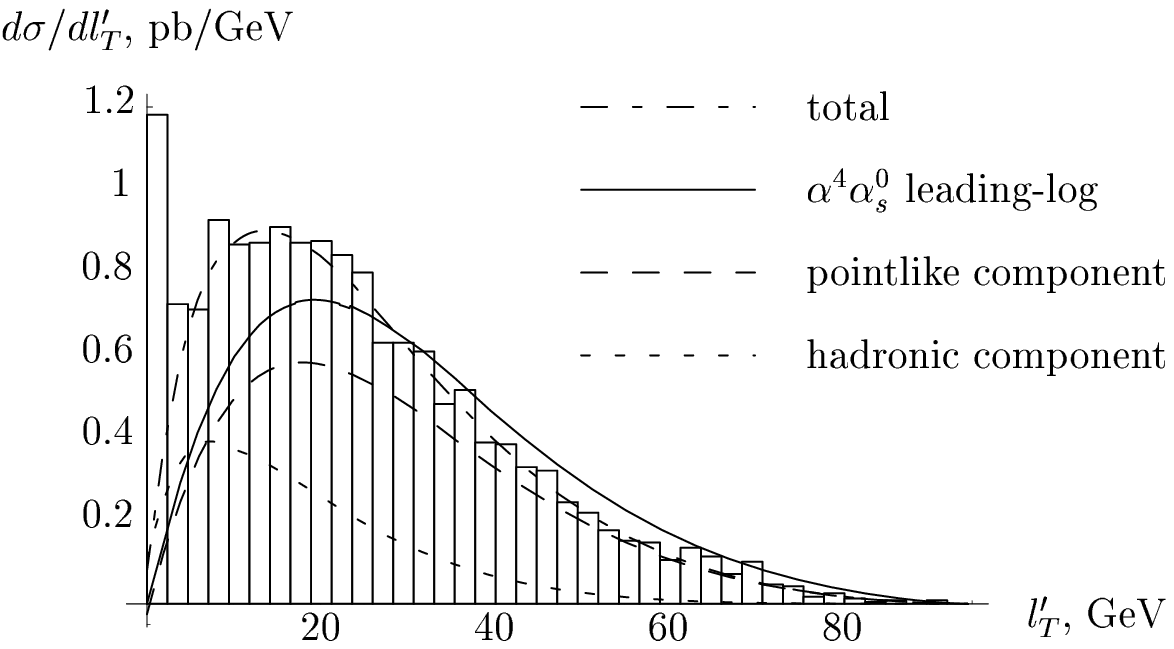}}
\vspace{0.2in}
\caption{Differential cross section $d\sigma/d l_T'$ versus transverse momentum of the neutrino. Our theoretical $\alpha^4 \alpha_s^0$ leading-log calculation and the contributions due to the pointlike component of the GRS parton distribution, the hadronic component of the GRS parton distribution and the total GRS parton distribution are superimposed on $\alpha^4 \alpha_s^0$ results from Monte Carlo simulations using GRC4F with 3666 events.}
\label{plotltGRS}
\end{figure}

We define the virtuality of the photon to be $- p^2$. In the
calculation that follows, we will make the approximation $p^2 =
-p_T^2$. The GRS parton distribution of a virtual photon has the form
\begin{eqnarray}
f_{q/\gamma}^{\mathrm{GRS}}(x,Q^2,-p^2) & = &
f_{q/\gamma}^{\mathrm{pl}}(x,Q^2,-p^2) +
f_{q/\gamma}^{\mathrm{had}}(x,Q^2,-p^2) \nonumber \\  & = & 
f_{q/\gamma}^{\mathrm{pl}}(x,Q^2,-p^2) \nonumber \\
& & +
\frac{\alpha}{(1-p^2/m_\rho^2)^2}\left( G_q^2 f_{q/\pi}(x,Q^2) +
\delta_q \frac{1}{2} (G_u^2 - G_d^2) f_{s/\pi}(x,Q^2)\right),
\end{eqnarray} 
with $G_u^2 \simeq 0.836$, $G_d^2 \simeq 0.250$, and $\delta_u=-1$,
$\delta_d=+1$. The two pieces of this parton distribution are the
pointlike piece $f_{q/\gamma}^{\mathrm{pl}}$ and the hadronic piece
$f_{q/\gamma}^{\mathrm{had}}$. The latter is a function of the parton
distribution in a pion $f_{q/\pi}$ and of the strange sea distribution
$f_{s/\pi}$, weighted by factors of order unity and a overall factor
$(1-p^2/m_\rho^2)^{-2}$ which turns on sharply as the virtuality $-p^2$
of the photon goes below the mass squared, $m_\rho^2$, of the $\rho$
meson. Here $\pi$ refers to the neutral pion $\pi^0$. The
nonperturbative hadronic part is seen to start playing a role in the
overall parton distribution at low photon virtuality, where the photon
can be na\"{\i}vely thought of as a vector meson. The distributions
$f_{q/\gamma}^{\mathrm{pl}}$, $f_{q/\pi}$ and $f_{s/\pi}$ are
specified parametrically by Gl\"uck, Reya, and Schienbein. We compute
the distribution of quarks in an electron using the convolution
\begin{eqnarray}
f_{q/e}^{\mathrm{GRS}}(x,Q^2) & = & \frac{\alpha}{2\pi}
\int_x^1\frac{d \xi}{\xi} \int_{m_e^2}^{M^2} \frac{d p_T^2}{p_T^2}
P_{\gamma/e}(x/\xi) f_{q/\gamma}^{\mathrm{GRS}}(\xi, Q^2, p_T^2 ) .
\end{eqnarray} 

\begin{figure}[ht]
\vspace{0.2in}
\centerline{\epsfbox{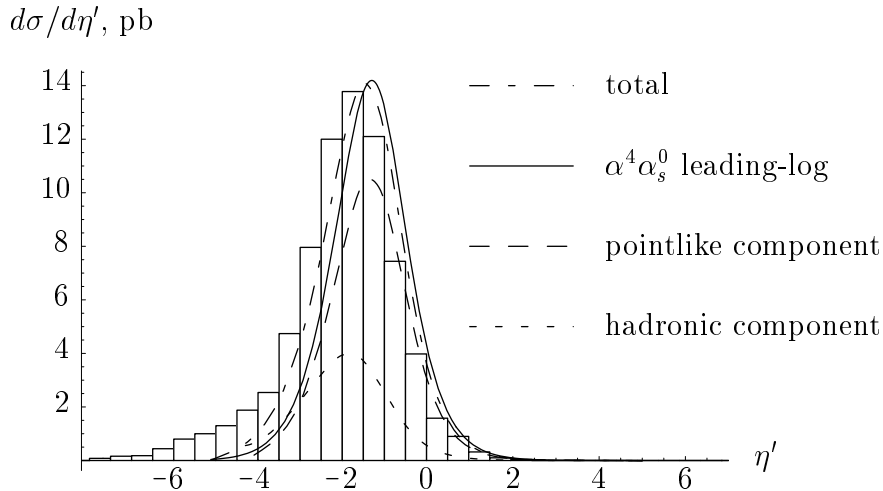}}
\vspace{0.2in}
\caption{Differential cross section $d\sigma/d \eta'$ versus pseudo-rapidity of the neutrino. Our theoretical $\alpha^4 \alpha_s^0$ leading-log calculation and the contributions due to the pointlike component of the GRS parton distribution, the hadronic component of the GRS parton distribution and the total GRS parton distribution are superimposed on $\alpha^4 \alpha_s^0$ results from Monte Carlo simulations using GRC4F with 3666 events.}
\label{plotetaGRS}
\end{figure}

We perform the calculations of the differential cross sections as
described earlier, using the new GRS distributions. The results are
displayed in Figs.~\ref{plotltGRS}~and~\ref{plotetaGRS}.

We notice a few interesting features of the differential cross
sections calculated using the GRS distributions. The cross section due
to the pointlike GRS distribution has the general shape of the cross
section we calculated using our unevolved pointlike distribution but
is somewhat smaller. We should mention that the lower cut-off on
$k_T$, $k_T^2 > \max( (0.28\ \mathrm{GeV})^2, p_T^2)$, we use to match
the GRC4F results is lower than the corresponding cut-off used by GRS,
$\max( 0.26\ \mathrm{GeV^2}, p_T^2 )$.  If we set our cut-off equal to
the GRS cut-off, the difference between our unevolved pointlike
distribution and the GRS pointlike distribution diminishes, but does
not vanish. One conjectures that this difference is due to the
evolution built into the parametrization of the GRS pointlike
distribution. The contribution due to the hadronic piece of the GRS
distribution is fairly soft, as one would expect. The differential
cross sections $d\sigma/d l_T'$ (Fig.~\ref{plotltGRS}) and $d\sigma/d
\eta'$ (Fig.~\ref{plotetaGRS}) are in reasonably close agreement with
the GRC4F Monte-Carlo data, except once again at low $l_T'$ and very
negative $\eta'$.

This check of our calculation and of the GRC4F Monte-Carlo leads us to
believe that the exact result of the differential cross sections
should lie reasonably close to the total GRS curves in
Fig.~\ref{plotltGRS}~and~\ref{plotetaGRS}, excluding perhaps the very
low $l_T'$ and very negative $\eta'$ regions. One estimate of the
error would be the difference between the GRS curves and the
$\alpha_s^0 \alpha_s^4$ leading-log curves.

We conclude that the GRC4F data is for the most part within the errors
of our calculation. This gives us confidence both in our calculations
and the GRC4F Monte Carlo. Finally, we have gained some insight into
the physical content of the different parts of the cross section.

\section*{Acknowledgments}
I thank Davison E. Soper for invaluable guidance and discussion. I
thank David Strom and Alain Bellerive of the OPAL collaboration for
suggesting this analysis and for their help with the GRC4F Monte Carlo
program. I thank the CERN Theory Division for its kind hospitality
during the period in which this calculation was carried out.

\end{document}